\documentclass[a4paper]{panle}
\usepackage{url}
\usepackage{cite}
\usepackage{wrapfig}
\usepackage{graphicx}
\usepackage{amsmath, amssymb, amsthm, amsfonts}
\usepackage{longtable}
\usepackage{rotating}
\usepackage{lscape}
\usepackage{epsfig}
\usepackage{multirow}
\usepackage{bm}
\usepackage{mathptmx}

\newtheorem{lemma}{Lemma}

\usepackage{lineno}

\originalTeX
\begin{document}

\title{Feshbach Resonances of the Second Order in a Friedrichs Model}
\maketitle
\authors{D.\,S.~Grudko\,$^{a,b,}$\footnote{E-mail: gds@theor.jinr.ru}, A.\,K.~Motovilov\,$^{a,b,}$\footnote{E-mail: motovilv@theor.jinr.ru}}
\setcounter{footnote}{0}
\from{$^{a}$\,Bogoliubov Laboratory of Theoretical Physics, JINR, Dubna, Russia}
\from{$^{b}$\,Dubna State University, Dubna, Russia}


\begin{abstract}

We consider an extended version of the matrix Friedrichs model that may produce Feshbach resonances of the second order. In this model, under enlarging of the coupling constant, a pair of eigenvalues of the unperturbed Hamiltonian embedded into the continuous spectrum first transforms in resonances of the first order. At some moment later on, these two resonances merge into a single resonance but of the second order, and we fix conditions for this event. Being based on the operator interpretation of resonances, we found that, in the case under consideration, the operator roots of the transfer function have a very particular structure.
\bigskip

\noindent\textit{\small To be published in Physics of Particles and Nuclei (PEPAN),\\ https://www.pleiades.online/en/journal/physpart/.}

\end{abstract}
\vspace*{6pt}

\noindent
PACS: 02.30.Tb; 03.65.$-$w; 03.65.Nk

\section*{INTRODUCTION}

The second-order resonance, that is, the one associated with a second-order pole of the scattering matrix is a much rarer phenomenon compared to the typically encountered resonances of the first order (see, e.g., \cite{DemDru:1966,Nussenzweig:1972,Newton:1982}
and references therein). However, being somewhat exotic, such a resonance is even more intriguing than a usual first-order resonance since it generates more sophisticated temporal decay of the corresponding resonant state involving powers of time \cite{GoldWatson:1964}.

It is well known that K.O.\,Friedrichs developed two celebrated model Hamiltonians \cite{Friedrichs:1948} in order to study perturbations of the continuous spectrum. One of the Hamiltonians \cite{Friedrichs:1948}, the matrix Friedrichs model is, in addition, well adjusted for easy understanding the nature of the Feshbach resonance. It is known that this model may generate a second-order resonance \cite{GadellaPronko:2011} but the latter can not be interpreted as a Feshbach one.

In the present work, we employ an extended version \cite{MennMot:1999} of the matrix Friedrichs model \cite{Friedrichs:1948}. We show that already a minimalist variant of the $2\times2$ Hamiltonian in \cite{MennMot:1999} with just two eigenvalues embedded into the continuous spectrum produces, under an appropriate perturbation, a resonance of the second order. Although of the second order, this resonance still remains a Feshbach resonance since it originates from a merger of two first-order explicitly Feshbach resonances.

\section{CONTINUED TRANSFER FUNCTION AND ITS OPERATOR ROOT}

We consider a $2\times2$ block operator matrix Hamiltonian of the form
\begin{equation}\label{Hamiltonian}
    {H}=\left(\begin{matrix}
        A_0 & B\\
        B^* & A_1
    \end{matrix}\right),
\end{equation}
acting in the orthogonal sum $\mathcal{H}=\mathcal{H}_0\oplus\mathcal{H}_1$ of the Hilbert spaces $\mathcal{H}_0=L_2(-1,1)$ and $\mathcal{H}_1=\mathbb{C}^2$. It is assumed that the entry $A_0$ is the multiplication by a variable, $(A_0u_0)(\mu) = \mu u_0(\mu)$, $\mu\in[-1,1]$, for any function $u_0\in \mathcal{H}_0$. The entry $A_1 = \text{diag}(-\lambda,\lambda)$ is a diagonal number matrix with $\lambda\in(0,1)$. The coupling $B:\mathcal{H}_1\to\mathcal{H}_0$ is defined by
\begin{equation}
    (Bu_1)(\mu)=B(\mu)u_1=b_1(\mu)u^{(1)}_1+b_2(\mu)u^{(2)}_1,
    \quad    u_1 = (u^{(1)}_1 ,u^{(2)}_1)^T\in\mathbb{C}^2,
\end{equation}
where $b_1,b_2\in\mathcal{H}_0$ and $B(\mu)$ is a row composed of the values $b_k(\mu)$, $\mu\in [-1,1]$. The entry $B^*$ is adjoint of $B$. Notice that the operator $H$ represents a version of the extended Friedrichs model in \cite{MennMot:1999}. Obviously, this operator is bounded and self-adjoint in $\mathcal{H}$.

All the singularities of the resolvent $({H}-z)^{-1}$ of $H$ different from those of the resolvent $({H}_0-z)^{-1}$ of the unperturbed Hamiltonian ${H}_0=\mathop{\rm diag}(A_0,A_1)$ are nothing but zeros of the \emph{transfer function}
\begin{equation}\label{SpecProblem}
    M_1(z):= A_1-z+V_1(z), \,\, \text{with} \,\,
    V_1(z):=-B^*(A_0-z)^{-1}B, \quad z\in\mathbb{C}\setminus[-1,1],
\end{equation}
that is, the points $z\in\mathbb{C}\setminus[-1,1]$ where the matrix $M_1(z)$ has the eigenvalue $0$ (see, e.g., \cite{MennMot:1999}). Furthermore, the resonances of $H$ are just the points $z\in\mathbb{C}$ (located already on the neighboring unphysical energy sheets) that are zeros for the analytic continuation of $M_1$ through the interval $(-1,1)$ (in case if such a continuation exists).

In the following we assume that $b_1(\mu)=b_2(\mu)=b=\text{const}$. We also suppose that $\Omega$ is a simply connected domain in the lower half-plane $\mathbb{C}^-$ of the complex plane $\mathbb{C}$ with piecewise-smooth boundary $\partial\Omega$ and that the interval $[-1,1]$ is a part of $\partial\Omega$. Let $\Gamma$ be the remaining part of $\partial\Omega$, that is, $\Gamma=\partial\Omega\setminus[-1,1]$. Under the above assumptions, analytic continuation of the transfer function $M_1$ from the upper half-plane $\mathbb{C}^+$ of $\mathbb{C}$ through the interval $(-1,1)$ into the part $\Omega$ of the lower half-plane $\mathbb{C}^-$ (already associated with an unphysical energy sheet) is explicitly given by
\begin{equation*}
    M_1(z,\Gamma):=A_1-z+V_1(z,\Gamma),\,\,\, \text{with}\,\,
    {V}_1(z,\Gamma):=\displaystyle|b|^2\left(\begin{matrix}
        1 & 1\\
        1 & 1
    \end{matrix}\right)\int\limits_\Gamma\dfrac{d\eta}{z-\eta}, \quad z\in\Omega.
\end{equation*}
Hence, the resonances of $H$ in the part $\Omega$ of the unphysical sheet involved are just the zeros of the function $M_1(\cdot,\Gamma)$.

Following \cite{MennMot:1999}, along with the function ${V}_1(z,\Gamma)$ of the scalar variable $z\in\mathbb{C}\setminus\Gamma$ we introduce a closely related transformation
\begin{equation}
    \label{V1G} {V}_1(Z,\Gamma):=\displaystyle|b|^2\left(\begin{matrix}
        1 & 1\\
        1 & 1
    \end{matrix}\right)\int\limits_\Gamma d\eta (Z-\eta)^{-1}, \quad
    \mathop{\rm spec}(Z)\cap\Gamma=\varnothing,
\end{equation}
that maps the set of $2\times 2$ matrices into itself. The next statement is a corollary of \cite[Theorem 3.1]{MennMot:1999}.

\begin{lemma}
    \label{lem} Assume that $|b|^2 < d_\Gamma^2/(8\ell_\Gamma)$ where $\ell_\Gamma$ denotes the length of the curve $\Gamma$ and $d_\Gamma$ stands for the distance between $\Gamma$ and the spectrum of $A_1$. Then the operator equation
    \begin{equation}\label{MasterEquation}
        {Z}={A}_1+{V}_1({Z},\Gamma)
    \end{equation}
    has a solution $Z\in\mathbb{C}^{2\times2}$ satisfying the bound $\|Z-A_1\|\leq\frac{d_\Gamma}{2}-\sqrt{\frac{d_\Gamma^2}{4}-2|b|^2\ell_\Gamma}$. The solution $Z$ is unique in the operator ball $\|Z-A_1\|<d_\Gamma-|b|\sqrt{2\ell_\Gamma}$ (and does not depend on $\Gamma$).
\end{lemma}

The transformation \eqref{V1G} possesses a remarkable property: If $u_1$ is an eigenvector of the argument matrix ${Z}$ corresponding to an eigenvalue $z$, that is, if $Zu_1=zu_1$, then ${V}_1({Z},\Gamma)u_1={V}_1(z,\Gamma)u_1$. Thus, if one applies both sides of the equation \eqref{MasterEquation} to an eigenvector $u_1$ of its solution $Z$, then the spectral identity $M_1(z,\Gamma)=0$ pops up. So, by solving \eqref{MasterEquation}, one gets a powerful tool for studying the spectrum of $M_1(\cdot,\Gamma)$ including the resonances of $H$. For a detail and more general exposition of the material presented we refer the reader to paper \cite{MennMot:1999}. Here we only notice that due to a factorization property \cite[Theorem 4.1]{MennMot:1999} the solution $Z$ of \eqref{MasterEquation} may be viewed as an operator root of the (continued) transfer function $M_1(z,\Gamma)$.

Under the assumptions adopted in the present paper, the unique solution $Z$ of \eqref{MasterEquation} has a quite particular structure.

\begin{lemma}
    \label{prop} The unique solution $Z$ to \eqref{MasterEquation} mentioned in Lemma \ref{lem} has the following form:
    \begin{equation}\label{Z-Matrix}
        {Z}=\left(\begin{matrix}
            -\alpha-\beta i & \gamma\\
            -\overline{\gamma} & \alpha-\beta i
        \end{matrix}\right),
    \end{equation}
    where $\alpha>0$, $\beta\geqslant0$, and $\gamma\in\mathbb{C}$ are such that $\|Z-A_1\|\leq\frac{d_\Gamma}{2}-\sqrt{\frac{d_\Gamma^2}{4}-2|b|^2\ell_\Gamma}$.%
\end{lemma}

Proof is done by an observation that the form \eqref{Z-Matrix} of the argument matrix is reproduced under iterations of \eqref{MasterEquation} and taking into account that the mapping on the r.h.s. of \eqref{MasterEquation} is a strict contraction (see \cite[Theorem 3.1]{MennMot:1999}).

\section{ANALYSIS OF THE TRANSFER FUNCTION}

Eigenvalues of a matrix of the form \eqref{Z-Matrix} read as
\begin{equation}
    \lambda_\pm = -\beta i\pm(\alpha^2-|\gamma|^2)^{1/2}.
\end{equation}
Both of them are simultaneously the zeros of the transfer function $M_1(\cdot,\Gamma)$ and there are no other zeros of $M_1(\cdot,\Gamma)$ in the $d_\Gamma/2$--neighborhood of the spectrum of $A_1$ (see \cite[Corollary 4.5]{MennMot:1999}). Clearly, the degeneration case corresponds to $|\gamma|=\alpha$ where the two eigenvalues merge into one sitting just on the imaginary axis. Thus, we make a substitution $z=-iy$, $y\geqslant 0$, and explicitly compute:
\begin{equation}\label{Determinant}
    \mathop{\rm det}{M}_1(-iy,\Gamma) = -\lambda^2 - y^2  +
    4|b|^2 y \left[\pi-\arctan{\left(\text{${1}/{y}$}\right)}\right].
\end{equation}
By using graph considerations we then prove that for any $\lambda\in(0,1)$ there always exist $b$ satisfying $\lambda^{1/2}(2\pi)^{-1/2}\leq |b|\leq\lambda^{1/2}(\pi+2\arctan\lambda)^{-1/2}$  such that the function $\mathop{\rm det}{M}_1(\cdot,\Gamma)$ possesses exactly one root, $z=-\beta i$, and this root is of the second order. It is just the root that generates the second-order pole for the inverse $M_1(\cdot,\Gamma)^{-1}$ of the analytically continued transfer function $M_1$ and, hence, it produces a second-order resonance for the operator $H$.

\section*{CONCLUSIONS}

This work shows that already a minimalist extension of the matrix Fried\-richs model with two embedded eigenvalues allows one to present a Hamiltonian possessing a Feshbach resonance of the second order. The operator root \eqref{Z-Matrix} of the transfer function $M_1(\cdot,\Gamma)$ has a very particular structure: it represents the sum of a $J$-self-adjoint matrix and a multiple of the identity, $-\beta i I$, which helps to clearly explain emergence of a second-order resonance by merging a couple of simple resonances.

\section*{CONFLICT OF INTEREST}

The authors declare that they have no conflicts of interest.

\bibliographystyle{pepan}
\bibliography{G03_biblio}

\end{document}